\documentclass[amsfonts, amssymb, amsmath, twocolumn, showkeys, nofootinbib, reprint, floatfix, prx, superscriptaddress]{revtex4-2}

\usepackage[colorinlistoftodos, color=green!40,prependcaption]{todonotes}
\usepackage{graphicx}% Include figure files
\usepackage{dcolumn}% Align table columns on decimal point
\usepackage{bm}% bold math
\usepackage{float}
\usepackage{gensymb}
\usepackage{xcolor}
\usepackage{amsmath}
\usepackage{amssymb}
\usepackage{units}
\usepackage{placeins}
\usepackage{ulem}
\usepackage{hyperref}
\usepackage{bbm}
\usepackage{multirow}
\usepackage{amsfonts}
\usepackage{amsmath}
\usepackage{amssymb}
\usepackage{xspace}
\usepackage[colorinlistoftodos]{todonotes}
\usepackage[T1]{fontenc}

\usepackage{lipsum}
\usepackage{comment}

\def\Neel{N\'{e}el\xspace}
\newcommand{\Lvec}{\ensuremath{\mathbf{L}}\xspace}

\begin{document}

\title{Strain continuously rotates the \Neel vector in altermagnetic MnTe}

\author{Alex Liebman-Pel\'{a}ez}
\author{Jon Kruppe}
\affiliation {Department of Physics, University of California, Berkeley, California 94720, USA}
\affiliation {Materials Science Division, Lawrence Berkeley National Laboratory, Berkeley, California 94720, USA}

\author{Resham Babu Regmi}
\author{Nirmal J. Ghimire}
\affiliation {Department of Physics \& Astronomy, University of Notre Dame, Notre Dame, Indiana 46556, USA}
\affiliation{Stavropoulos Center for Complex Quantum Matter, University of Notre Dame, Notre Dame, IN, USA 46556}

\author{Yue Sun}
\affiliation{Department of Physics, University of Washington, Seattle, Washington 98195, USA}

\author{Igor I. Mazin}
\affiliation{Department of Physics and Astronomy, George Mason University, Fairfax,
VA 22030, USA}
\affiliation{Quantum Science and Engineering Center, George Mason University, Fairfax, VA 22030, USA}

\author{Hilary M. L. Noad}
\affiliation{Max Planck Institute for Chemical Physics of Solids, 01187 Dresden, Germany}

\author{James Analytis}
\affiliation {Department of Physics, University of California, Berkeley, California 94720, USA}
\affiliation {Materials Science Division, Lawrence Berkeley National Laboratory, Berkeley, California 94720, USA}

\author{Veronika Sunko}
\affiliation {Institute of Science and Technology Austria, Vienna, Austria}

\author{Joseph Orenstein}
\email{jworenstein@lbl.gov}
\affiliation {Department of Physics, University of California, Berkeley, California 94720, USA}
\affiliation {Materials Science Division, Lawrence Berkeley National Laboratory, Berkeley, California 94720, USA}

\begin{abstract}
Altermagnetism has recently emerged as a distinct class of collinear antiferromagnets that break time-reversal symmetry, exhibiting a host of novel properties. Applied strain has attracted particular attention as a key tuning parameter for altermagnets. Although several experimental studies have demonstrated the preparation of single-domain states through a combination of applied strain and magnetic field, the route to such states remains unclear. Here, we use magneto-optical measurements on single crystals of MnTe under applied strain to show that, in contrast to previous reports, strain acts primarily to rotate the \Neel vector \Lvec continuously. Since the orientation of \Lvec determines the magnetic point group symmetry, this continuous rotation effectively tunes the symmetry and its associated physical properties. Furthermore, we demonstrate that built-in strain in free-standing crystals is sufficient to pin \Lvec into continuous textures over millimeter length scales. Together, these results provide guidance for future device design and open the door to leveraging the \Neel vector orientation as a tunable degree of freedom in spintronic applications.
\end{abstract}

\maketitle

\begin{figure*}[!htbp]
    \centering
    \includegraphics[scale=1]{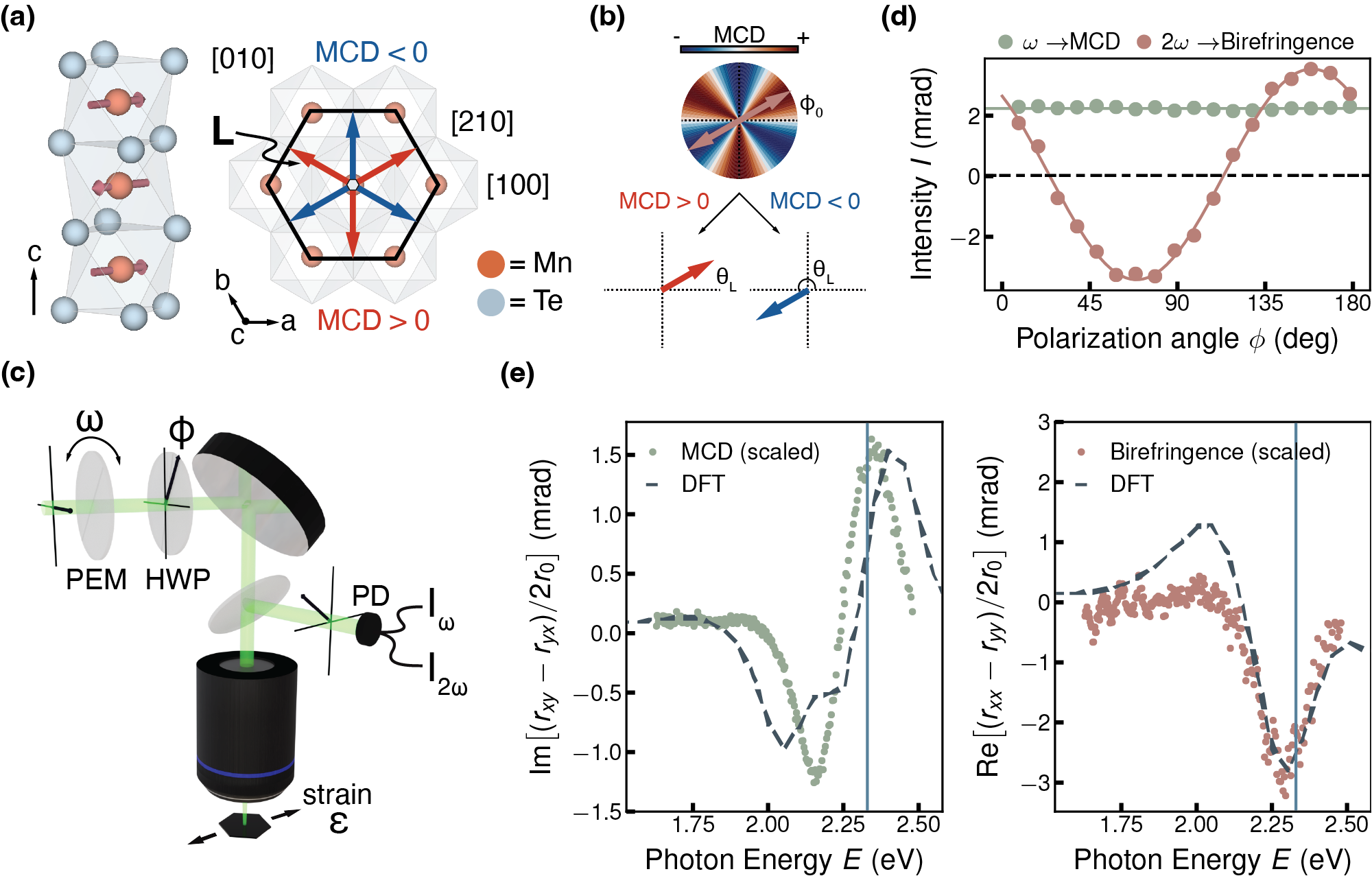}
    \caption{\textbf{(a)} Crystal structure of MnTe. The MCD is maximized when \Neel\ vector \Lvec points along next-nearest-neighbor Mn-Mn bonds, with the sign flipping under $C_{6z}$ rotations. \textbf{(b)} Protocol for determining the \Neel vector orientation $\theta_L$ from birefringence angle $\phi_0$ and the sign of MCD. \textbf{(c)} Optical setup for measuring MCD and birefringence by PEM-modulated reflectivity at $\omega$ and $2\omega$ for incident polarization $\phi$ The incident polarization is set by a half-wave plate (HWP). \textbf{(d)} Normalized reflected intensity of $\unit[2.33]{eV}$ light in $\omega$ and $2\omega$ channels, indicating clear MCD and birefringence in MnTe. \textbf{(e)} MCD and birefringence optical spectra compared with ab-initio DFT calculations for MnTe with \Lvec oriented along $[210]$. Blue vertical line indicates $\unit[2.33]{eV}$.}
    \label{fig:setup}
\end{figure*}

%\linenumbers
\section{Introduction}

Recently, a class of antiferromagnets, known as altermagnets, has been recognized to host large spin-splitting of bands that far exceeds the relativistic (spin-orbit) contribution \cite{sarma_spintronics_2004, jungwirth_antiferromagnetic_2016, baltz_antiferromagnetic_2018, smejkal_crystal_2020, gonzalez-hernandez_efficient_2021, smejkal_beyond_2022, smejkal_anomalous_2022, fender_altermagnetism_2025, jungwirth_altermagnetic_2025}. The intense interest in altermagnets stems from their capacity to host spin-polarized currents while remaining insensitive to stray magnetic fields. To implement device strategies based on this property, it is essential to understand how to manipulate the \Neel vector \Lvec, which serves as the primary order parameter \cite{mcclarty_landau_2024}.

Applied strain is potentially a powerful tool to manipulate altermagnetic order \cite{khodas_tuning_2026}. Strain is predicted to drive electronic transitions \cite{karetta_strain-controlled_2025}, and piezomagnetic effects \cite{aoyama_piezomagnetic_2024} can lead to, for example, the realization of tunable random-field Ising models \cite{chakraborty_magnetic-field-tuned_2025}. In hexagonal $\alpha$-MnTe, the prototypical $g$-wave altermagnet \cite{wasscher_evidence_1965, gonzalez_betancourt_spontaneous_2023, lovesey_templates_2023, kluczyk_coexistence_2024, krempasky_altermagnetic_2024, lee_broken_2024}, theoretical and experimental work has demonstrated that strain can change the symmetry of the spin-split Fermi surface \cite{belashchenko_giant_2025}, enhance non-relativistic spin splittings \cite{chen_strain_2026}, prepare single domain states \cite{amin_nanoscale_2024, liu_strain-tunable_2025}, and tune the strength and sign of the anomalous Hall effect (AHE) \cite{smolenski_strain-tunability_2025, liu_strain-tunable_2025}. However, an understanding of how strain affects the \Neel vector remains incomplete.

Here, we report spatially-resolved optical measurements on single crystals of MnTe under tunable, in-situ strain. The applied strain is seen to reorient the \Neel vector through continuous rotation, rather than via the domain detwinning mechanism described in Ref.~\cite{liu_strain-tunable_2025}. Since the orientation of \Lvec determines the magnetic point group symmetry, continuous rotation effectively tunes the symmetry and its associated physical properties --- for instance, allowing the anomalous Hall effect to be ``turned off." By applying larger strains, we observe hysteresis of \Lvec, suggesting that the magnetic subsystem can undergo a magnetic analog to plastic deformation. The dominance of strain over magnetocrystalline anisotropy (MCA) in setting the orientation of \Lvec suggests that uncontrolled strains, such as those built-in during crystal growth, affect \Lvec. We confirm this by imaging an unstrained sample, finding that built-in strain pins \Lvec into smooth textures spanning a wide range of orientations over millimeter length scales. 

%Our results thus provide practical insights into how to manipulate the \Neel vector in MnTe, and open the door to utilizing a previously hidden degree of freedom for device design: the continuous orientation of \Lvec.

\begin{figure*}[!htbp]
    \centering
    \includegraphics[scale=1]{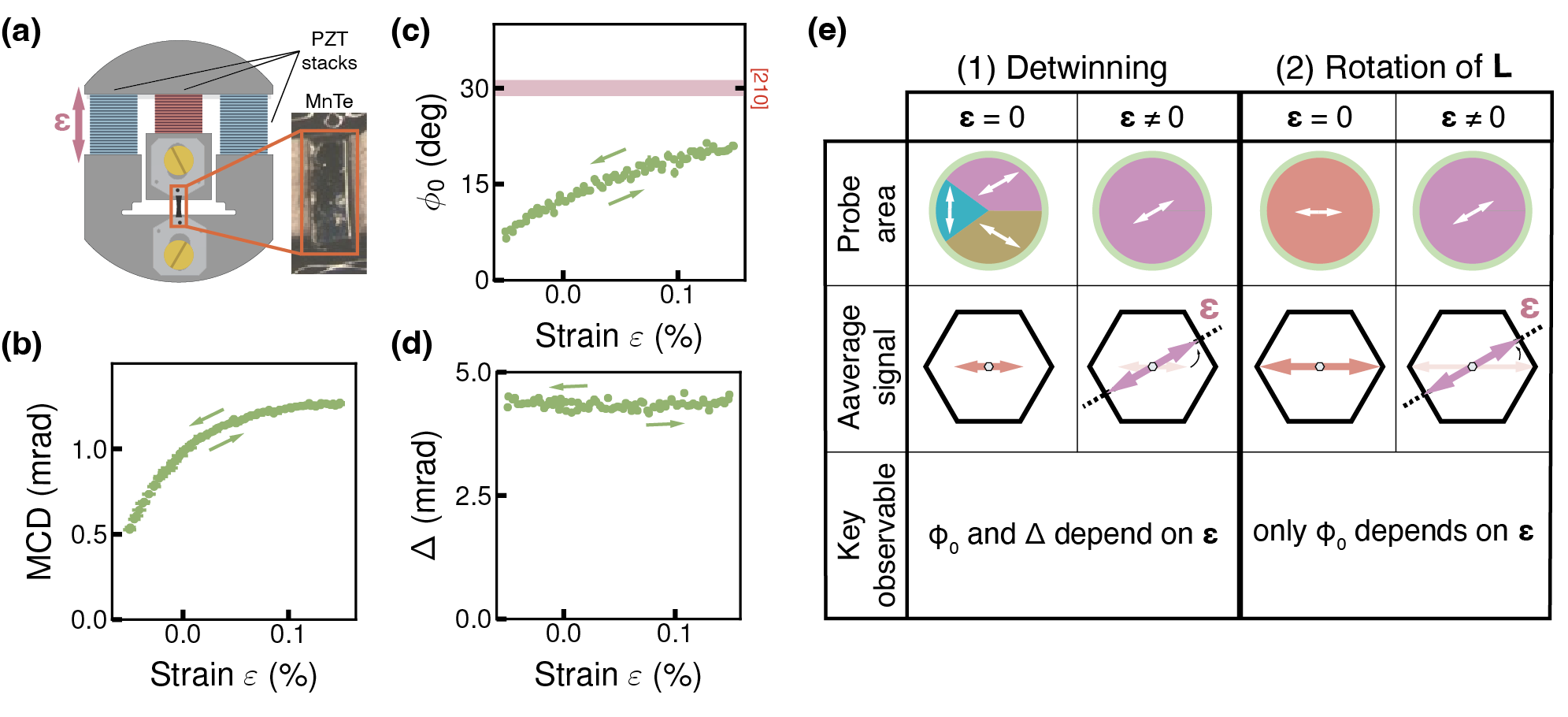}
    \caption{\textbf{(a)} Illustration of a MnTe sample mounted into a piezoelectric strain cell. Strain $\varepsilon$ is applied along either the $[100]$ or $[210]$ directions, and optical measurements are performed at $\unit[25]{K}$. Strain $\varepsilon$ evolution of \textbf{(b)} MCD, \textbf{(c)} $\phi_0$, and \textbf{(d)} $\Delta$ for a sample S1 cooled under zero strain conditions. The pink band in (c) indicates that strain was applied along $[210]$, and arrows show that strain was swept in both directions. The strain-independence of $\Delta$ indicates that strain rotates \Lvec continuously. \textbf{(e)} Comparison of the effect of strain on the birefringence signal for two models: (1) detwinning, and (2) rotation of \Lvec. For (1), in general both $\phi_0$ and $\Delta$ shift with strain as domain populations under the probe area change with strain. On the other hand, for (2) only $\phi_0$ changes. The strain evolution shown here corresponds to compressive strain applied along $[210]$.}
    \label{fig:strain_1}
\end{figure*}

\section{Optical response of the \Neel vector in MnTe}

$\alpha$-MnTe is a magnetic material composed of hexagonal layers of Mn stacked along the $\mathbf{c}$ direction, where each Mn is in octahedral coordination with Te. The two Mn$^{2+}$ ions carry moments $\mathbf{M}_1$ and $\mathbf{M}_2$ ($S=5/2$, $L=0$) that order below $T_N = \unit[307]{K}$, as shown in Fig.~\ref{fig:setup}a \cite{kunitomi_neutron_1964}. Moments within each magnetic sublattice align ferromagnetically in the $\mathbf{ab}$ plane, and the sublattices align antiparallel to each other. As such, the primary order parameter in MnTe is the \Neel vector $\mathbf{L} = \mathbf{M}_1 - \mathbf{M}_2$. In a single domain, the two sublattices are related by a six-fold screw rotation of the ions combined with time reversal, a symmetry that produces $g$-wave altermagnetism \cite{smejkal_emerging_2022}. Furthermore, the hexagonal symmetry suggests that the spins are subject to a six-fold MCA, and therefore exhibit six domains corresponding to spins along the $\langle210\rangle$ family \cite{kriegner_magnetic_2017, liu_strain-tunable_2025}. These are the symmetry-aligned \Neel vector domains that have been proposed to detwin by applied strain \cite{liu_strain-tunable_2025}.

We are able to probe \Lvec because it modifies the $2\times2$ reflectivity tensor $r_{ij}$, which determines the reflection amplitude of light at normal incidence, in two distinct ways. First, since \Lvec lies in the $\mathbf{ab}$ plane, it breaks the three-fold rotational symmetry of the paramagnetic state and induces birefringence ($r_{xx}\neq r_{yy})$. \Lvec lies along one of the two principal optical axes, whose orientation we denote as the birefringence angle $\phi_0$. However, measurement of $\phi_0$ leaves ambiguity as to the orientation of \Lvec along that axis.

To resolve the direction of \Lvec, we measure magnetic circular dichroism (MCD), which is the optical analog of the AHE. MCD is associated with the antisymmetric off-diagonal component of the reflectivity tensor, $r_{xy} = - r_{yx}$ and obeys the same symmetry constraints as $\sigma_{xy} = - \sigma_{yx}$. The magnitude and sign of MCD is sensitive to the orientation of \Lvec \cite{mcclarty_landau_2024, amin_nanoscale_2024, liu_strain-tunable_2025, gonzalez_betancourt_spontaneous_2023, kriegner_magnetic_2017}. For instance, when \Lvec points along the next-nearest-neighbor Mn-Mn bond directions $\langle210\rangle$, the magnetic point group (MPG) is $m^\prime m^\prime m$ and MCD is symmetry-allowed \cite{lovesey_templates_2023, aoyama_piezomagnetic_2024}. On the other hand, when \Lvec points along the nearest-neighbor Mn-Mn bond directions $\langle100\rangle$, the MPG is $mmm$ and MCD is symmetry-forbidden. More generally, the contribution of the \Neel vector to MCD can be expanded to first order as,
\begin{equation}
    \textnormal{MCD} \propto  L^3\sin(3\theta_L),
    \label{eq:MCD_vs_L}
\end{equation}
where $\mathbf{L} = L(\cos\theta_L, \sin\theta_L)$, $L\geq0$, and the \Neel vector orientation $\theta_L$ is measured with respect to the $[10\bar{1}0]$ direction \cite{mcclarty_landau_2024}. Notice that while birefringence is invariant under rotation of \Lvec by $\unit[180]{\degree}$, which is equivalent to time-reversal, the sign of the MCD flips. As illustrated in Fig.~\ref{fig:setup}b, the sign of MCD can therefore be used in combination with the birefringence angle $\phi_0$ to pin down $\theta_L$ \cite{amin_nanoscale_2024}.

\subsection{Optical Methods}\label{sec:optical_methods}

A schematic of the optical setup for measuring birefringence and MCD is shown in Fig.~\ref{fig:setup}c. A linearly polarized probe beam passes through a photoelastic modulator (PEM) that modulates the light between linear and circular polarization at a frequency $\omega/2\pi \approx \unit[50]{kHz}$. The beam is then focused onto the sample through an objective at normal incidence. The reflected intensity $I$ is collected on a photodiode, demodulated at both $\omega$ and $2\omega$, and normalized to the DC reflected intensity. The signal at $\omega$ is independent of the incident polarization $\phi$ and is given by
\begin{equation}
    I_\omega = \pi J_1\left(\frac{\pi}{2}\right)\textnormal{MCD},
\end{equation}
where $\textnormal{MCD} = \textnormal{Im}[(r_{xy} - r_{yx})/2r_0]$,  $r_0 = (r_{xx} + r_{yy})/2$, and $J_i(z)$ is the $i^\textnormal{th}$ Bessel function of the first kind. The signal at $2\omega$ is given by
\begin{equation}
    I_{2\omega}(\phi) = \pi J_2\left(\frac{\pi}{2}\right)\Delta\cos2(\phi + \phi_0),
    \label{eq:MLD}
\end{equation}
where $\Delta = \textnormal{Re}[(r_{xx} - r_{yy})/2r_0]$ is the birefringence amplitude and $\phi_0$ is the birefringence angle with respect to the $[100]$ direction. Further details about the experimental setup can be found in Supplemental Material Sec.~\ref{sec:SM_optics}. Fig.~\ref{fig:setup}d shows an example of both signals in MnTe at $T=\unit[25]{K}$ for a photon energy of $\unit[2.33]{eV}$.

%We note that a shift of $\unit{-12.5}{\degree}$ has been applied to $\phi_0$ for all measurements to account for a systematic misalignment of the setup.  

\subsection{\Neel vector vs weak ferromagnetism in the optical response}\label{sec:spectroscopy}

Because MCD and a weak ferromagnetic moment $\mathbf{M} = \mathbf{M}_1 + \mathbf{M}_2$ \cite{mazin_origin_2024} have the same symmetry, there is a question as to whether the MCD arises from \Lvec or $\mathbf{M}$. When MCD is probed by X-rays resonant at the $L_{2,3}$ edges, the response has been demonstrated to originate from \Lvec \cite{hariki_x-ray_2024, amin_nanoscale_2024}.  Extending the same spectroscopic approach to the visible regime, we demonstrate that MCD directly probes the sign of $\Lvec$. In Fig.~\ref{fig:setup}e we show MCD and birefringence spectra of MnTe for light in the range $\unit[1.5-2.5]{eV}$, together with theoretical predictions that assume $\mathbf{M}=0$ (see Supplemental Material Sec.~\ref{sec:SM_dft}). The close agreement between experiment and theory indicates that both MCD and birefringence signals arise from \Lvec, demonstrating the validity of our method for probing its magnitude and direction.

% Now that we understand the relation between \Lvec and the optical response, we turn to investigating how applied strain affects \Lvec.

\begin{figure*}[!htbp]
    \centering\
    \includegraphics[scale=1]{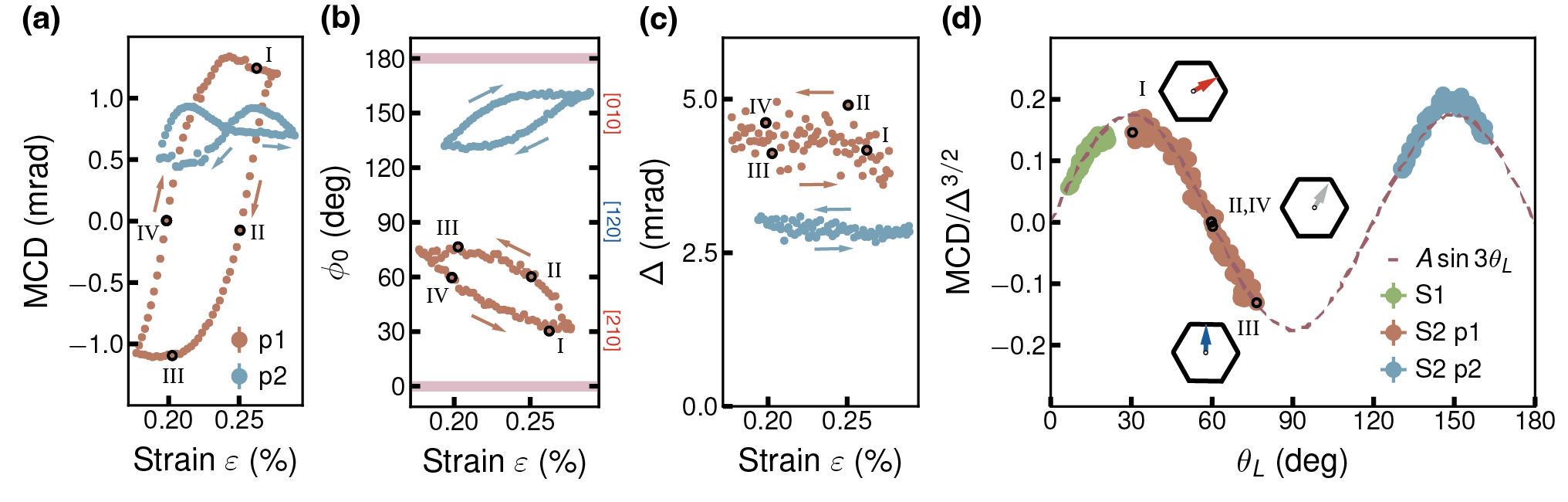}
    \caption{Strain evolution of \textbf{(a)} MCD, \textbf{(b)} birefringence angle $\phi_0$, and \textbf{(c)} birefringence amplitude $\Delta$ for a sample S2 cooled under the differential thermal strain between the strain cell and the crystal, and with strain applied along $[100]$ (pink bands in panel (b)). Measurements are taken at two different positions on the sample p1 and p2. Strain hysteresis suggests plasticity of the magnetic subsystem. \textbf{(d)} Correlation between $\textnormal{MCD}/\Delta^{3/2}$ and \Neel vector orientation $\theta_L$ for S1, p1, and p2. Collapse onto Eq.~\eqref{eq:MCD_vs_L} confirms a picture where strain continuously rotates \Lvec. Least-squares fit to data yields $A = 0.174$. At p1, MCD is tuned to zero when \Lvec aligns with $[110]/\unit{60}{\degree}$. Cartoons of \Lvec and the corresponding MCD signal are depicted for p1 at various points along the hysteresis loop $\textnormal{I}\rightarrow\textnormal{II}\rightarrow\textnormal{III}\rightarrow\textnormal{IV}$.}
    \label{fig:strain_2}
\end{figure*}

\section{Applied strain dependence of the \Neel vector}

To directly track the strain evolution of the \Neel vector, we applied in-situ strain to high-quality single crystals of MnTe \cite{jost_chiral_2025} and measured both the magnitude and direction of \Lvec using the methods described above. In Fig.~\ref{fig:strain_1}a, we show a schematic of the applied strain experiment. Samples were cut into bars and loaded into a Razorbill CS130 piezoelectric strain cell. The sample was heated above $T_N$ and then cooled to $\unit[25]{K}$. We tuned the strain by applying voltages to the PZT stacks and monitoring displacements on the sample with a capacitive displacement sensor (see Supplemental Material Sec.~\ref{sec:SM_strain} for more details). Because of the differential thermal expansion between MnTe and the titanium cell, at $\unit[25]{K}$ the crystal can be under considerable tensile strain in the absence of applied strain. This is similar to epitaxial strain in that strain is applied during the cooling process. However, using the piezo stacks, compressive strain can be applied during cooling to maintain nominally zero strain on the sample.

% We used protocols---compensating and not compensating for thermal strain---to verify the validity of our conclusions across different straining protocols.

In one experiment, we mounted sample S1 ($\unit[0.7]{mm}\times\unit[0.26]{mm}\times\unit[0.2]{mm}$) with $[210]$ along the strain axis and cooled it under zero strain. In Fig.~\ref{fig:strain_1}b--d, we show the evolution of MCD, $\phi_0$, and $\Delta$ as a function of strain $\varepsilon$ at $\unit[25]{K}$. The key observation is that, while strain tunes MCD and $\phi_0$ continuously, $\Delta$ is independent of strain. As we discuss below, and illustrate in Fig.~\ref{fig:strain_1}e, this feature of the data is consistent with a continuous rotation of the \Neel vector and inconsistent with detwinning.

The detwinning picture assumes that the signal sensed by the $\approx\unit[5]{\mu m}$ probe beam corresponds to the area-weighted average signal of domains with \Lvec aligned along the $\langle210\rangle$ family \cite{hwangbo_strain_2024}. Since birefringence cannot distinguish between time-reversed states, these six domains give rise to three birefringent domains for which $\phi_0 = \unit{30}{\degree}, \unit{90}{\degree}$, and $\unit{150}{\degree}$. Depending on the relative areas, the averaged orientation of \Lvec, encoded in $\phi_0$, can assume any value. However, as $\phi_0$ varies through the high-symmetry directions, the averaged magnitude of \Lvec, encoded in $\Delta$, must change as well. In Fig.~\ref{fig:strain_1}e we show an example where a mixed domain state is detwinned by strain to select for a single domain. The averaged value of $\Delta$ is necessarily enhanced in the detwinned state relative to the mixed state. A full analysis of the variation in $\Delta$ that accompanies tuning of $\phi_0$ in the detwinning model is presented in Supplemental Material Sec.~\ref{sec:SM_microdomain}. 

The detwinning picture is clearly inconsistent with the data presented in Fig.~\ref{fig:strain_1}b--d, which show that as strain continuously rotates \Lvec and changes MCD within the probe volume, $\Delta$ remains constant. The continuous reorientation of \Lvec with strain indicates that the magnetoelastic (ME) contribution to the free energy dominates the magnetocrystalline (MCA) term, as observed in other hexagonal magnetic systems \cite{donoway_multimodal_2024}. This is also consistent with the isotropic easy-plane observed in electron spin resonance measurements \cite{povarov_low-energy_2025}. In this regime, the orientation of \Lvec is set by the competition between applied strain and other sources of strain, such as strain built-in during crystal growth. This can explain why at $\varepsilon=0$, \Lvec lies along $\unit[13]{\degree}$, which is not aligned to an MCA easy-axis. We discuss the behavior of \Lvec and $\phi_0$ as a function of the relative strengths of ME and MCA in Supplemental Material Sec.~\ref{sec:SM_strainmodel}.

\begin{figure*}[!htbp]
    \centering
    \includegraphics[scale=1]{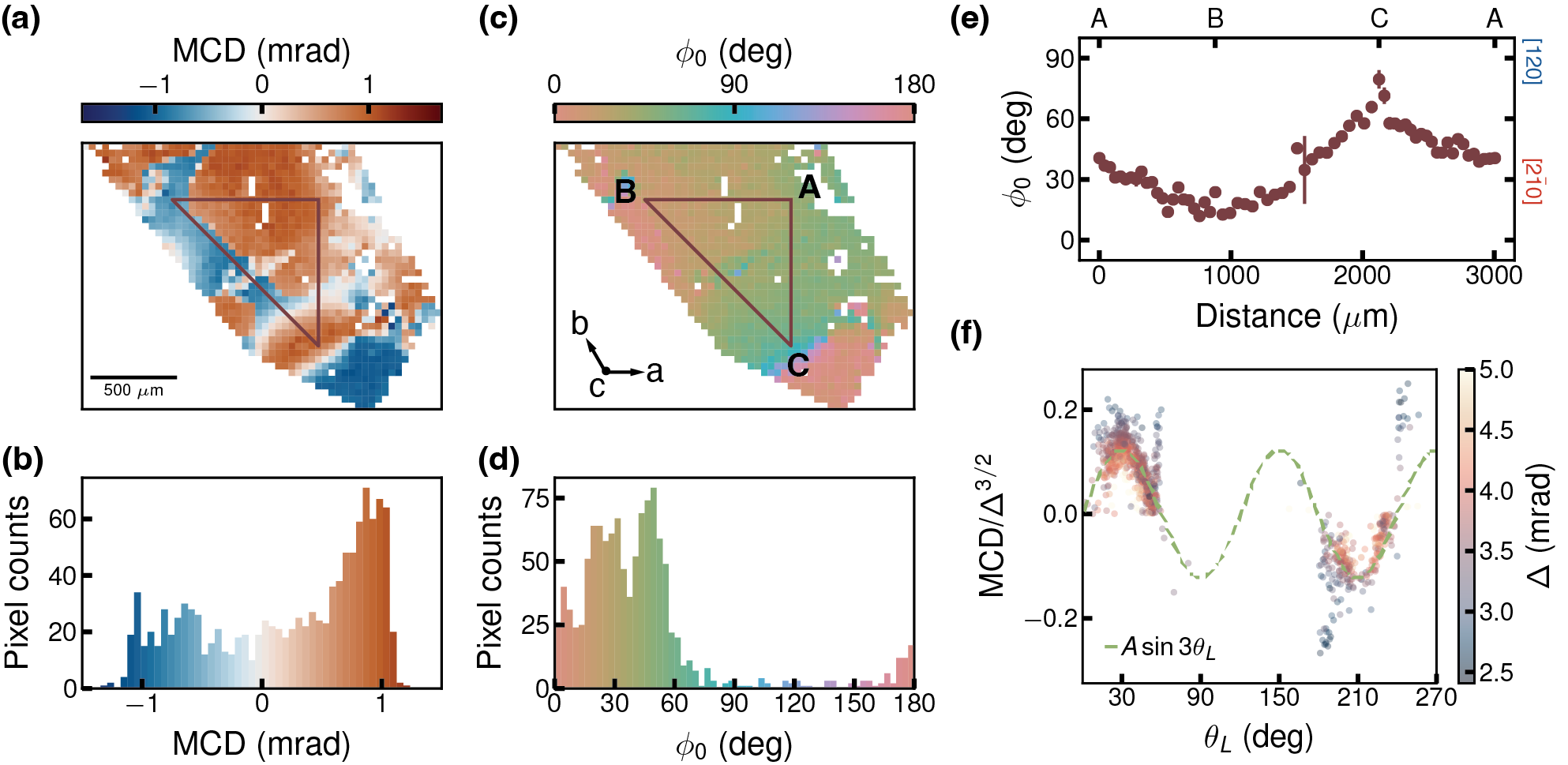}
    \caption{Maps and histograms of \textbf{(a-b)} MCD, and \textbf{(c-d)} $\phi_0$ across a large free sample at $\unit[25]{K}$, showing continuously varying $\phi_0$. \textbf{(e)} $\phi_0$ as a function of distance around the closed loop A$\rightarrow$B$\rightarrow$C$\rightarrow$A shown in (a) and (c). Vertical lines indicated error bars. \textbf{(f)} Collapse of $\textnormal{MCD}/\Delta^{3/2}$ versus $\theta_L$ across the free sample onto Eq.~\eqref{eq:MCD_vs_L}. Points are colored according to the birefringence amplitude $\Delta$, and are made partially transparent to highlight high-density regions. Least-squares fit to data yields $A = 0.123$.}
    \label{fig:mapping}
\end{figure*}

Next, we present the results of an experiment in which sample S2 ($\unit[2]{mm}\times\unit[0.7]{mm}\times\unit[0.2]{mm}$) is cooled without compensating for the thermal strain. The crystal was oriented with the $[100]$ direction along the strain axis. At $\unit[25]{K}$, the sample was under a tensile thermal strain of $\unit[0.23]{\%}$, and applied strain was used to tune around that point. Fig.~\ref{fig:strain_2}a--c shows the strain evolution of MCD, $\phi_0$, and $\Delta$ at two positions, p1 and p2. Under this larger strain, both MCD and $\phi_0$ have become hysteretic functions of strain at both positions. The difference in response at the two positions likely arises from the difference in the local built-in strain. The shape of the hysteresis loops is reminiscent of stress-strain hysteresis observed in crystals undergoing plastic deformation \cite{dowling_mechanical_2012}. However, we verified that the mechanical response of a MnTe crystal under comparable conditions is in the elastic regime (see Supplemental Material Sec.~\ref{sec:SM_elastic}), indicating that the hysteresis arises from the magnetic subsystem. Despite these differences between samples cooled under thermal strain and zero strain, the key observation remains the same: strain tunes MCD and $\phi_0$ but not $\Delta$. This indicates that, regardless of the straining protocol, applied strain primarily rotates the \Neel vector continuously.

The ability to rotate \Lvec continuously by applying strain enables us to test the predicted sinusoidal dependence of MCD on $\theta_L$ expressed by Eq.~\eqref{eq:MCD_vs_L}. First, we determine $\theta_L$ by measurement of $\phi_0$ and the sign of MCD. As $\Delta \propto L^2$, we take $\Delta^{3/2}$ to be proportional to the $L^3$ prefactor in Eq.~\eqref{eq:MCD_vs_L}. In Fig.~\ref{fig:strain_2}d, we show the dependence of $\textnormal{MCD}/\Delta^{3/2}$ on $\theta_L$ for samples S1 and S2 as strain sweeps the values of both. The collapse of all measurements on a single curve proportional to $\sin(3\theta_L)$ (i) confirms the prediction of Eq.~\eqref{eq:MCD_vs_L}, and (ii) confirms that strain primarily acts to rotate $\theta_L$, irrespective of the cooling protocol.

In addition, Fig.~\ref{fig:strain_2} highlights the hysteresis loop at p1 as it proceeds through configurations labeled by Roman numerals from $\textnormal{I}\rightarrow\textnormal{IV}$. As strain rotates \Lvec from $\theta_L\approx\unit{30}{\degree}$ to $\unit{90}{\degree}$, the sign of the MCD flips, passing through zero at $60^{\degree}$ (configuration II,IV).   

\section{Spatial dependence of the \Neel vector in unstrained MnTe}\label{sec:mapping}

Above, we posited that applied strain dominates over the MCA to set the orientation of the \Neel vector in MnTe. We also suggested that even in the absence of applied strain, \Lvec can be pinned by other sources of strain, such as that built-in during crystal growth. We now turn to spatial mapping of \Lvec across a large, free-standing single crystal to verify this assertion.

In Fig.~\ref{fig:mapping}a--d we show maps and histograms of MCD and $\phi_0$ for a single crystal of MnTe at $\unit[25]{K}$. The sample was mounted onto a copper plate with a thin layer of Apiezon N grease for thermal contact and rastered under the beam to produce the maps. We observe smooth variations of $\phi_0$ over a range of approximately $\unit{70}{\degree}$ across the millimeter-sized sample, as shown in the linecuts in Fig.~\ref{fig:mapping}e. Furthermore, $\Delta$ is constant across the majority of the sample, indicating that the variation in $\phi_0$ primarily reflects variations in the \Neel vector orientation rather than averaging over symmetry-locked domains (Fig.~\ref{fig:SI_histogram}). In Fig.~\ref{fig:mapping}f we show the collapse of $\textnormal{MCD}/\Delta^{3/2}$ onto Eq.~\eqref{eq:MCD_vs_L} across the whole sample, similar to the case of deliberately applied strain. Points are colored according to their respective values of $\Delta$, and we display only those where the uncertainty in $\phi_0$ is smaller than $\unit{3}{\degree}$, representing $\unit[80]{\%}$ of all measured points. Most points lie near the sinusoidal curve, consistent with $\theta_L$ varying spatially due to a continuous texture in the built-in strain. Points that deviate from Eq.~\eqref{eq:MCD_vs_L} tend to have small $\Delta$, suggesting that in these regions $\theta_L$ varies on length scales smaller than the probe beam size and averaging effects become prominent. Taken together, we conclude that built-in strain distributions are sufficient to pin \Lvec over a continuous range of angles across macroscopic length scales. We note that the remaining $\unit[20]{\%}$ of points not displayed here are in the bottom right corner of the sample, where $\Delta$ is even smaller and points tend to lie far off the sinusoidal curve.

\section{Conclusions}

In this work, we combined optical imaging with in-situ strain to demonstrate that, in contrast to previous reports on MnTe \cite{liu_strain-tunable_2025}, strain primarily rotates the \Neel vector rather than detwinning degenerate domains. This finding holds regardless of whether the crystal is cooled under zero-strain conditions or subject to thermal strains. In the latter case, hysteresis in the \Neel vector orientation was observed on cycling additional applied strain. We further demonstrated that strains built into a free-standing crystal are sufficient to pin the \Neel vector over a continuum of orientations over millimeter length scales. The observation of magnetic textures over such long length scales complements other results on single crystal showing nanoscale texture in both magnetic and electronic sectors \cite{yamamoto_altermagnetic_2025, ma_nanoscale_2026}, and adds to the growing literature cataloging sources of sample-to-sample variations in bulk properties \cite{zhou_surface-state-driven_2026, bey_conductivity_2026}. 

%Since neutron scattering measures the average properties of a bulk crystal, it may also explain the discrepancy with the previous report of strain detwinning \cite{liu_strain-tunable_2025}. 

We also reported that, for a certain positions on the crystal, strain induced a reversal of the sign of MCD. Similar strain-induced sign reversals of the AHE have been previously reported \cite{smolenski_strain-tunability_2025, liu_strain-tunable_2025}, and proposed to originate from distortions of the multipolar Berry curvature for a fixed \Neel vector direction. In contrast, here we find that a change in sign of MCD can be understood as a rotation of \Lvec. These can be compatible results; the low-frequency and optical responses to strain need not be the same, and in Refs.\cite{smolenski_strain-tunability_2025, liu_strain-tunable_2025} the sign of the AHE was determined with a magnetic field whereas here we measure MCD at zero magnetic field. However, the \Neel vector has not been tracked simultaneously in either measurement of the AHE. This motivates further study of the interplay between Berry curvature effects and the orientation of \Lvec in determining the strain response of the AHE. In addition, the role of piezomagnetic effects in determining the sign of the AHE should be revisited with the understanding that strain can rotate \Lvec. Spatially-resolved optical experiments that combine in-situ strain and applied magnetic fields will be particularly insightful in establishing the relationship between the strain dependence of the AHE and MCD.

Further study of the strain hysteresis is an opportunity for future research as well. The similarity between the hysteresis we observed in MCD versus strain and the hysteresis associated with stress versus strain suggests a magnetic analog of plasticity, confined to the spin system alone. Just as plastic deformation involves the motion and pinning of dislocations, spin system hysteresis could be mediated by irreversible rearrangement of magnetic textures. The threshold between elastic and plastic behavior, and the ways in which spin ``memory" might be useful in spintronic applications are unexplored. Since the plastic deformation regime likely corresponds to the high strains achievable by epitaxial strain, understanding spin plasticity, in combination with the insight that strain tunes the orientation of \Lvec, may be of relevance to device design and fabrication.

The continuum of \Lvec orientations, each with distinct physical properties, represents a previously unrecognized tuning knob for altermagnetic devices. At the same time, our spatial mapping demonstrates that uncontrolled strains arising during crystal growth or thin film deposition can impose unwanted variations in the orientation of \Lvec across a device. Managing these strain landscapes will be an important consideration for realizing altermagnetism-based devices.

\begin{acknowledgments}
This research was primarily funded by the Quantum Materials (KC2202) program under the U.S. Department of Energy, Office of Science, Office of Basic Energy Sciences, Materials Sciences and Engineering Division under Contract No. DE-AC02-05CH11231, which supported the experimental and theoretical work at the LBNL and UC Berkeley. N.J.G., R.~B.~R., and I.I.M. were supported by Army Research Office under Cooperative Agreement Number W911NF- 22-2-0173. H.M.L.N. and V.S. acknowledge funding through the Deutsche Forschungsgemeinschaft (DFG, German Research Foundation) through Grant No. TRR288—422213477, Project No. A10. H.M.L.N. acknowledges financial support from the Max Planck Society. Research in Dresden benefits from the environment provided by the DFG Cluster of Excellence ctd.qmat (EXC2147, Project ID 390858490).
\end{acknowledgments}

%\clearpage
\bibstyle{apsrev4-1}
\bibliography{references}

% reset things for SI$
\counterwithout{equation}{section}
\renewcommand\theequation{S\arabic{equation}}
\renewcommand\thefigure{S\arabic{figure}}
\renewcommand\thetable{S\arabic{table}}
\renewcommand\thesection{S\arabic{section}}
\renewcommand\bibnumfmt[1]{[S#1]}
\setcounter{equation}{0}
\setcounter{figure}{0}
\setcounter{enumiv}{0}

\clearpage
\newpage

\section*{Supplemental Materials}

\section{Optical technique}\label{sec:SM_optics}

Here we derive the correspondence $I_\omega\rightarrow \textnormal{MCD}$ and $I_{2\omega}\rightarrow \textnormal{birefringence}$. The polarization state of light passing through the optical system depicted in Fig.~\ref{fig:setup}c can be tracked using the Jones calculus. In addition to the components shown in Fig.~\ref{fig:setup}c, an optical chopper at frequency $\omega_c = \unit[533]{Hz}$ is placed along the beam path before the half-wave plate (HWP).

For light propagating at normal incidence, the polarization state in the plane normal to the propagation direction is given by a complex vector $\mathbf{E}$. Light enters the system with linear horizontal polarization so that the initial state is
\begin{equation}
    \mathbf{E}_0 = \begin{pmatrix}
        1 \\ 0
    \end{pmatrix}.
\end{equation}
The effect of each optical component on the polarization state is given by a Jones matrix. Horizontal and vertical polarizations corresponds to $y$ and $x$ on the sample, respectively. For a quarter-wave PEM oriented with its optic axis at a $\unit{45}{\degree}$ angle relative to the vertical axis, the Jones matrix in the linear polarized basis is
\begin{equation}
    J_\textnormal{PEM}(t) =\begin{pmatrix}\cos\left[\frac{\pi}{4}\sin\omega t\right] & i\sin\left[\frac{\pi}{4}\sin\omega t\right] \\ i\sin\left[\frac{\pi}{4}\sin\omega t\right] & \cos\left[\frac{\pi}{4}\sin\omega t\right]\end{pmatrix}
\end{equation}
where $t$ is time and $\omega=\unit[50]{kHz}$. The PEM modulates the light polarization between linear and circular at angular frequency $\omega$. The HWP oriented with its optic axis at an angle $\phi/2$ has Jones matrix
\begin{equation}
    J_\textnormal{HWP}(\phi) = \begin{pmatrix}
        \sin\phi & \cos\phi \\ \cos\phi & -\sin\phi
    \end{pmatrix}.
\end{equation}
The HWP is used to set the time-averaged polarization $\phi$ incident on the sample, as indicated in Fig.~\ref{fig:setup}c. Incident vertical polarization, which corresponds to light along the $[10\bar{1}0]$ direction on the sample, is achieved for $\phi = 0$. Lastly, the Jones matrix for the sample is the reflectivity tensor $r_{ij}$,
\begin{align}
    &J_\textnormal{MnTe} = \begin{pmatrix}
        r_{xx} & r_{xy} \\ r_{yx} & r_{yy}
    \end{pmatrix} \\
    &= r_0\left[\mathbf{1} + R^{-1}\left(\phi_0\right) \begin{pmatrix}
        \tilde{\Delta} & 0 \\ 0 & -\tilde{\Delta}
    \end{pmatrix}R\left(\phi_0\right) +  \begin{pmatrix}
            0 & \tilde{\gamma} \\ -\tilde{\gamma} & 0
        \end{pmatrix}\right],\label{eq:SI_reflectivity}
\end{align}
where $R(\theta) = \begin{pmatrix}
    \sin\theta & -\cos\theta \\ \cos\theta & \sin\theta
\end{pmatrix}$ is a rotation of theta relative to the vertical axis. Here, $r_0 = (r_{xx} + r_{yy})/2$ is the bare reflectivity, $\tilde{\Delta} = (r_{x^\prime x^\prime} - r_{y^\prime y^\prime})/2r_0$ is the complex birefringence (primed coordinates have been rotated by the birefringence angle $\phi_0$ referenced to $x$), and $\tilde{\gamma} = (r_{xy} - r_{yx})/2r_0$ contains the combined MCD and magneto-optical Kerr effects.

The reflected electric field is given by
\begin{equation}
    \mathbf{E}(\phi, t) = J_\textnormal{MnTe}J_\textnormal{HWP}(\phi)J_\textnormal{PEM}(t)\mathbf{E}_0,
\end{equation}
and the measured intensity on the photodiode is
\begin{equation}
    I(\phi, t) = |E(\phi, t)|^2\frac{1}{2}\left(1 + \textnormal{sgn}(\sin\omega_ct)\right),
\end{equation}
where the square wave term accounts for the modulation by the optical chopper and $|E(\phi, t)|$ is the absolute value of the reflected electric field. Keeping only the first term in the Fourier transform of the square wave, we have
\begin{equation}
    I(\phi, t) = |E(\phi, t)|^2\left(\frac{1}{2} + \frac{2}{\pi}\sin\omega_ct\right).
    \label{eq:SI_signal}
\end{equation}
Using the Jacobi-Anger expansions, we expand $|E(\phi, t)^2|$ to first order in $\tilde{\Delta}$ and $\tilde{\gamma}$. Keeping only leading terms at DC, $\omega$, and $2\omega$, we obtain
\begin{equation}
    |E(\phi, t)|^2 \approx r_0^2  + E^2_\omega(t) + E^2_{2\omega}(\phi,t).
\end{equation}
Here,
\begin{align}
    E^2_\omega(t) &= 4r_0^2J_1\left(\frac{\pi}{2}\right)\textnormal{Im}\left[\tilde{\gamma}\right]\sin\omega t \\
    E^2_{2\omega}(\phi, t) &= 4r_0^2J_2\left(\frac{\pi}{2}\right)\textnormal{Re}\left[\tilde{\Delta}\right]\cos2(\phi + \phi_0)\cos2\omega t,
\end{align}
and $J_l(z)$ are the Bessel functions of the first kind. We simultaneously demodulate the measured intensity at $\omega_c$, $\omega$, and $2\omega$ using a lock-in amplifier to obtain the three RMS signals
\begin{align}
    I_{\omega_c}^{RMS} &= \frac{2}{\sqrt{2}}\frac{r_0^2}{\pi}\\
    I_\omega^{RMS} &= \frac{2}{\sqrt{2}}r_0^2J_1\left(\frac{\pi}{2}\right)\textnormal{Im}\left[\tilde{\gamma}\right]\\
    I_{2\omega}^{RMS} &= \frac{2}{\sqrt{2}}r_0^2J_2\left(\frac{\pi}{2}\right)\textnormal{Re}\left[\tilde{\Delta}\right]\cos2(\phi + \phi_0).
\end{align}
The point is that $I_{\omega_c}^{RMS}$ contains information about the diagonal reflectivity $r_0$, whereas $I_{\omega}^{RMS}$ and $I_{2\omega}^{RMS}$ contain information about MCD and birefringence respectively. Dividing the latter by $I_{\omega_c}^{RMS}$ yield the expressions for the normalized signals
\begin{align}
I_\omega &= \frac{I_\omega^{RMS}}{I_{\omega_c}^{RMS}} = \pi J_1\left(\frac{\pi}{2}\right)\textnormal{Im}\left[\tilde{\gamma}\right]\\
&\approx 1.78073\times\textnormal{Im}\left[\tilde{\gamma}\right]\\
I_{2\omega} &= \frac{I_{2\omega}^{RMS}}{I_{\omega_c}^{RMS}} = \pi J_2\left(\frac{\pi}{2}\right)\textnormal{Re}\left[\tilde{\Delta}\right]\cos2(\phi + \phi_0)\\
&\approx 0.78446\times\textnormal{Re}\left[\tilde{\Delta}\right]\cos2(\phi + \phi_0).
\end{align}
In practice, both signals are subject to a background, which is measured at $T > T_N$ and subtracted from the data before any further analysis.

\section{First-principles calculation of optical response in MnTe}\label{sec:SM_dft}

First-principles calculations were performed within DFT using the Vienna ab
initio simulation package (VASP) \cite{kresse_efficient_1996}. Generalized gradient approximation (GGA) \cite{perdew_generalized_1996} functional was used. On-site electronic correlations in Mn 3d states were
taken into account using the DFT+U method, with an effective Hubbard parameter
$U_{eff}=U-J=$4 eV. Mn pseudotential included the semicore $p$ states (Mn\_pv
pseudopotential). $k$-point mesh included up to $26\times26\times24$
divisions. The plane-wave cutoff was $\unit[400]{eV}$ and the number of bands included
in the calculation was 130.

\section{Applied strain experiments and strain determination}\label{sec:SM_strain}

\begin{figure*}
    \centering
    \includegraphics[scale=1]{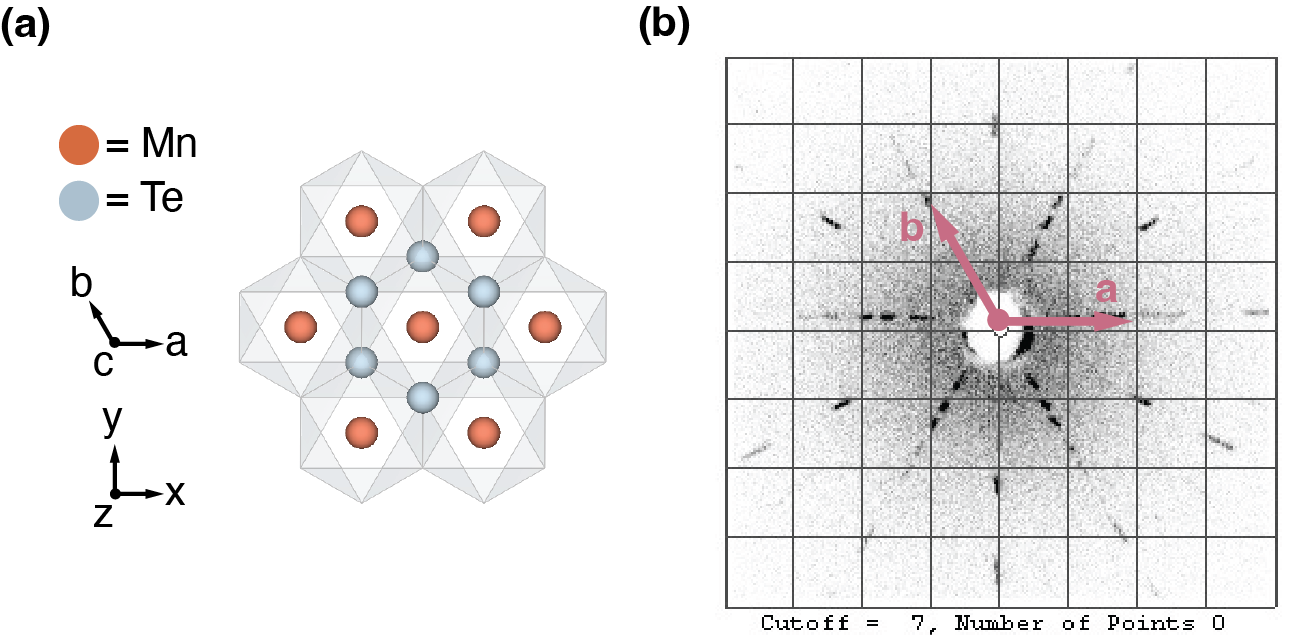}
    \caption{\textbf{(a)} Comparison between primitive cell lattice vectors $\mathbf{a}, \mathbf{b}, \mathbf{c}$ used in the main text and conventional cell lattice vectors $\mathbf{x}, \mathbf{y}, \mathbf{z}$ used in elasticity tensor $C_{ijkl}$. \textbf{(b)} Laue diffraction pattern used to align MnTe crystals.}
    \label{fig:SI_straindetermination}
\end{figure*}

In-situ strain was applied to single crystals of MnTe using a commercial three stack piezo-actuated strain cell (Razorbill CS130). Crystals were aligned using the Laue diffraction method, cut into bars using either a wire saw or a laser mill, and then loaded into the strain cell (Fig.~\ref{fig:SI_straindetermination}b-c). The crystal is suspended over a gap, with each end glued between thin titanium plates using STYCAST-2850FT/CAT9. The titanium plates are screwed into the cell on each side of the gap.

Uniaxial stress is applied to the sample by applying a voltage to the three piezo stacks using a Razorbill RP100, which effectively pushes/pulls the crystal at the ends. A capacitive displacement sensor, monitored with a Keysight
E4980AL LCR meter, is used to measure the strain along the stack direction. The stress-strain relation is given by $\sigma_{ij} = C_{ijkl}\varepsilon_{kl}$, where $\sigma_{ij}$ is the stress tensor, $\varepsilon_{kl}$ is the strain tensor, and $C_{ijkl}$ is the elasticity tensor. The elasticity tensor for MnTe has been calculated by DFT and can be found on the Materials Project \cite{de_jong_charting_2015}. Fig.~\ref{fig:SI_straindetermination}a shows a comparison between the primitive lattice vectors $\mathbf{a}, \mathbf{b}, \mathbf{c}$, and the conventional cell lattice vectors $\mathbf{x}, \mathbf{y}, \mathbf{z}$ associated with the elasticity tensor. In Voigt notation, the stress-strain relation is
\begin{equation}
    \left(\begin{matrix}\sigma_{xx}\\\sigma_{yy}\\\sigma_{zz}\\\sigma_{yz}\\\sigma_{xz}\\\sigma_{xy}\end{matrix}\right) = \left(\begin{matrix} 82&37&32&0&0&0\\37&82&32&0&0&0\\32&32&60&0&0&0\\0&0&0&28&0&0\\0&0&0&0&28&0\\0&0&0&0&0&28\\\end{matrix}\right) \left(\begin{matrix}\varepsilon_{xx}\\\varepsilon_{yy}\\\varepsilon_{zz}\\2\varepsilon_{yz}\\2\varepsilon_{xz}\\2\varepsilon_{xy}\end{matrix}\right),
    \label{eq:SI_stress_strain}
\end{equation}
where $C_{ijkl}$ is in units of GPa.

We measured two samples with applied stress along two different directions. For S1, which corresponds to Fig.~\ref{fig:strain_1},  stress was applied along $\mathbf{y}$/$\langle210\rangle$\footnote{Strain along $\mathbf{y}$ is equivalent to strain along $\unit[30]{\degree}$, since both corresponds to the $\langle210\rangle$ family.}. In this case the only non-zero element of the stress tensor is $\sigma_{yy}$, and the displacement sensor measures $\varepsilon_{yy}$. Using Eq.~\eqref{eq:SI_stress_strain}, the total strain on the sample is given by $\varepsilon_{yy}$, $\varepsilon_{xx} = -0.3\varepsilon_{yy}$ and $\varepsilon_{zz} = -0.37\varepsilon_{yy}$. For S2, stress was applied along $\mathbf{x}$/$\langle100\rangle$. In this case, only $\sigma_{xx}\neq0$, the displacement sensor measures $\varepsilon_{xx}$, and the total strain on the sample is specified by $\varepsilon_{xx}$, $\varepsilon_{yy} = -0.3\varepsilon_{xx}$, and $\varepsilon_{zz}=-0.37\varepsilon_{xx}$. In all cases, $\varepsilon>0$ is taken as tensile strain, and $\varepsilon<0$ is compressive.

Focusing on the case of stress applied along $\mathbf{x}$, the strain on the sample is determined by
\begin{equation}
    \varepsilon_{xx} = \eta\left(\varepsilon^V_{xx} + \varepsilon^T_{xx}\right),
\end{equation}
where $\varepsilon_{xx}^V$ is the piezo-actuated applied strain, $\varepsilon_{xx}^T$ is the thermal strain due to the differential thermal expansion of the cell and the sample, and $\eta$ is a transmission factor that accounts for the finite stiffness of the epoxy between the titanium plates and the sample. Previous finite element calculations have shown that for a comparable setup $\eta\approx 0.7$ \cite{liu_strain-tunable_2025}. It is notoriously difficult to determine the zero-strain point in strain cells equipped with only a strain gauge; we outline our approximations below.

The displacement of the sample plates $\Delta L$ is obtained by measuring the capacitance $C$ of the strain gauge
\begin{equation}
    \Delta L = \frac{\alpha}{C -C_a} -d_0,
\end{equation}
where $\alpha$, $C_a$, and $d_0$ are supplied by Razorbill. $\Delta L$ as measured above may be expressed as
\begin{equation}
    \Delta L = \Delta L^V + \Delta L^T + \Delta L^s,
\end{equation}
where $\Delta L^V$ is the voltage-dependent displacement of the peizo stacks, $\Delta L^T$ is the temperature-dependent displacements of the strain cell gap, and $\Delta L^s$ is the temperature-dependent displacements of the sensor. $\Delta L^s$ does not contribute to the strain. The large thermal expansion of the piezo stacks is compensated in the three stack geometry.

At a fixed temperature, we isolate the piezo-actuated applied displacement $\Delta L^V$ by subtracting $\Delta L^0 = \Delta L^T + \Delta L^s$, the measured displacement at 0 applied voltage, such that
\begin{equation}
    \varepsilon_{xx}^V = \frac{\Delta L^V}{L_\textnormal{MnTe}} \approx \frac{\Delta L^V}{L^0_\textnormal{MnTe}},
\end{equation}
where $L_\textnormal{MnTe}$ is the temperature-dependent length of MnTe, which is approximately equal to the length measured at room temperature $L^0_\textnormal{MnTe}$.

\begin{figure}
    \centering
    \includegraphics[scale=1]{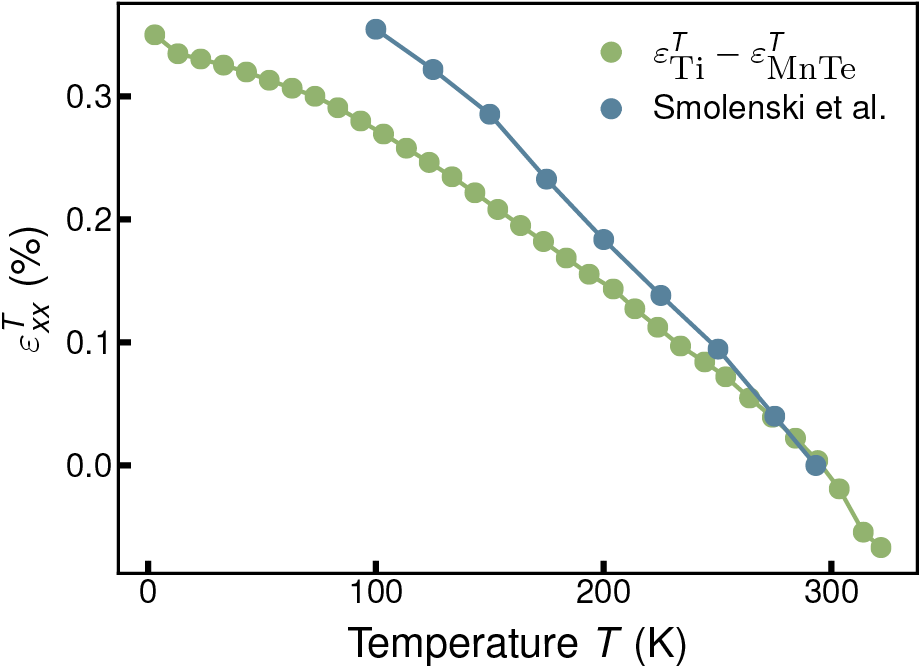}
    \caption{Differential thermal expansion of titanium and MnTe compared to $\varepsilon_{xx}^T$ as evaluated in Ref.~\cite{smolenski_strain-tunability_2025}.}
    \label{fig:SI_diffthermexp}
\end{figure}

Noting that the gap and the sample have the same length at room temperature in zero-strain conditions, the thermal strain is given by
\begin{equation}
    \varepsilon_{xx}^T = \frac{L_\textnormal{gap} - L_\textnormal{MnTe}}{L_\textnormal{MnTe}} = \frac{\Delta L^T - \Delta L^T_\textnormal{MnTe}}{L_\textnormal{MnTe}},
\end{equation}
where $L_\textnormal{gap}$ is temperature-dependent length of the gap and $\Delta L^T_\textnormal{MnTe}$ is the thermal expansion of free MnTe. Since isolating $\Delta L^T$ is challenging and nevertheless prone to systematic errors, we approximate it to be equal to the thermal expansion of titanium over the length of the gap. Therefore,
\begin{equation}
    \varepsilon_{xx}^T \approx \varepsilon_\textnormal{Ti}^T - \varepsilon_\textnormal{MnTe}^T,
\end{equation}
where $\varepsilon_\textnormal{Ti}^T$ and $\varepsilon_\textnormal{MnTe}^T$ are the relative thermal expansion for titanium and MnTe respectively \cite{corruccini_thermal_1961, aoyama_piezomagnetic_2024}. As shown in Fig.~\ref{fig:SI_diffthermexp}, this approximation is fairly accurate compared to $\varepsilon_{xx}^T$ as evaluated more carefully in Ref.~\cite{smolenski_strain-tunability_2025} for a similar strain cell. In summary, the strains displayed in the text are given by
\begin{equation}
    \varepsilon_{xx} = 0.7\left(\frac{\Delta L^V}{L^0_\textnormal{MnTe}} + \varepsilon_\textnormal{Ti}^T - \varepsilon_\textnormal{MnTe}^T\right).
\end{equation}
Similar consideration are used to evaluate $\varepsilon_{yy}$ for stress applied along $\mathbf{y}$.

\section{Detwinning model}\label{sec:SM_microdomain}

\begin{figure}
    \centering
    \includegraphics[scale=1]{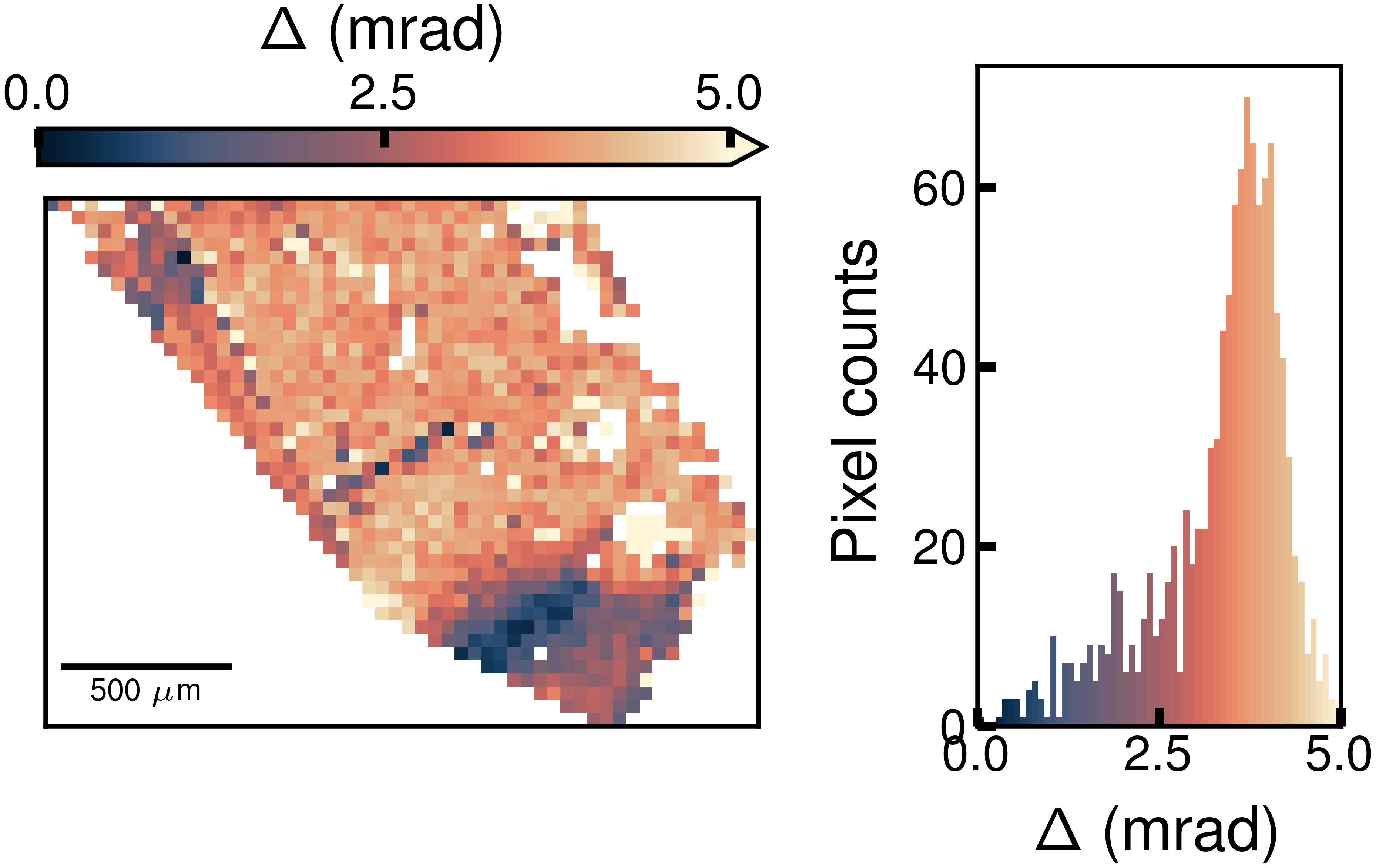}
    \caption{Map and histogram for birefringence amplitude $\Delta$ across the free sample associated with the spatial mapping shown in Fig.~\ref{fig:mapping}a-d.}
    \label{fig:SI_histogram}
\end{figure}

\begin{figure}
    \centering
    \includegraphics[scale=1]{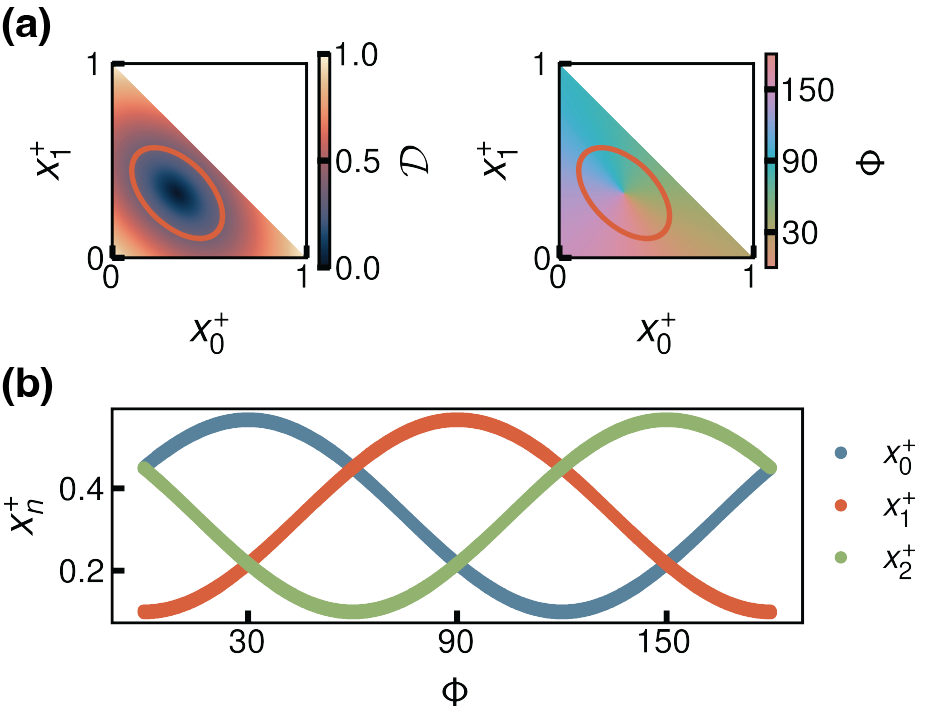}
    \caption{\textbf{(a)} The effective birefringence amplitude $\mathcal{D}$ and angle $\Phi$ over the $\mathbf{x}^+$ phase space. A trajectory is shown for which $\Phi$ changes continuously while $\mathcal{D}=0.5$. Here, $\Delta=1$. \textbf{(b)} The values of $x_0^+$, $x_1^+$, and $x_2^+$ along the trajectory indicated in (a).}
    \label{fig:SI_microdomains}
\end{figure}

In Fig.~\ref{fig:mapping} and Fig.~\ref{fig:SI_histogram} we saw that $\phi_0$ varies continuously across a sample of MnTe, in contrast to the expectation that the histogram for $\phi_0$ be peaked at $\unit{30}{\degree}$, $\unit{90}{\degree}$, and $\unit{150}{\degree}$, corresponding to domains with \Lvec aligned to the $\langle210\rangle$ family. In addition, in Fig.~\ref{fig:strain_1} we saw that strain induces a change in $\phi_0$ but not in the birefringence amplitude $\Delta$. For both free and strained crystals, the relation between MCD and $\theta_L$ (reconstructed from MCD and $\phi_0$) follows Eq.~\eqref{eq:MCD_vs_L}. Here, we explore the possibility that these observations can be accounted for within the detwinning model.

In the detwinning model, we consider a distribution of small domains under the laser spot. The possible orientations for \Lvec in each domain are  $\theta_L = \theta_n = \unit{30}{\degree} + n\unit{60}{\degree}$ for $n=0\rightarrow5$. We denote the population fraction of each domain with $\theta_n$ as $x_n$, such that $\sum_n x_n$ = 1. The reflectivity tensor for a domain with \Lvec at orientation $\theta_n$ can be obtained from Eq.~\eqref{eq:SI_reflectivity} with $\phi_0 = \theta_n$ and $\textnormal{sign}\left[\tilde{\gamma}\right] = (-1)^n$. Here, we take $\Delta = \textnormal{Re}\left[\tilde{\Delta}\right]$ as the birefringence amplitude and $\gamma = \textnormal{Im}\left[\tilde{\gamma}\right]$ as the MCD.

As discussed in Sec.~\ref{sec:SM_optics}, each domain contributes to both $I_\omega$ (MCD) and $I_{2\omega}$ (birefringence). The total signals in the detwinning scenario are thus proportional to
\begin{align}
    I_\omega &= \gamma\sum_n x_n(-1)^n\\
    I_{2\omega} &= \sum_n x_n\Delta\cos2(\phi + \theta_n),
\end{align}
where $\phi$ is the incident polarization. We now define $x_0^\pm = x_0 \pm x_3$, $x_1^\pm = x_1 \pm x_4$ and $x_2^\pm = x_2 \pm x_5$. These signals can then be rewritten as functions of $\mathbf{x^+} = (x_0^+,x_1^+,x_2^+)$ and $\mathbf{x^-} = (x_0^-,x_1^-,x_2^-)$,
\begin{align}
    I_\omega(\mathbf{x}^-) &= \gamma(x^-_0 - x_1^- + x^-_2)\\
    I_{2\omega}(\mathbf{x}^+) &= \mathcal{D}(\mathbf{x}^+)\cos2(\phi + \Phi(\mathbf{x}^+)),
\end{align}
where the effective birefringence amplitude $\mathcal{D}(\mathbf{x}^+)$ and principle axis direction $\Phi(\mathbf{x}^+)$ are given by
\begin{align}
 \mathcal{D}(\mathbf{x}^+) &= \frac{\Delta}{2} \sqrt{3(x^+_0 - x^+_2)^2 + (x^+_0 - 2x^+_1 + x^+_2)^2} \label{eq:microdomain_amp} \\
 \Phi(\mathbf{x}^+) &= \frac{1}{2}\tan^{-1}\left( \frac{\sqrt{3}(x^+_0 - x^+_2)}{x^+_0 - 2x^+_1 + x^+_2} \right). \label{eq:microdomain_ang}
\end{align}
The MCD is only non-zero if there is an imbalance between populations of opposite \Lvec ($x_n^-$), and the birefringence is only sensitive to the total number of domains along a given axis ($x_n^+$).

Having developed a model for both signals within the detwinning scenario, we now turn to evaluating how the various observations might be reconciled within this model. We start with the observation that strain causes $\phi_0$ to rotate without changing $\Delta$. We look for what kind of strain induced changes to $\mathbf{x} = (x_0,...,x_5)$ are necessary to match this observation. The normalization constraint $\sum_n x_n = 1$ allows us to rewrite Eqs.\ \eqref{eq:microdomain_amp} and \eqref{eq:microdomain_ang} in terms of just two components of $\mathbf{x}^+$, arbitrarily taken as $x^+_0$ and $x^+_1$. Fig.~\ref{fig:SI_microdomains}a shows the effective birefringence amplitude and principle axis within this reduced parameter space. As indicated by the ellipses, we find that the only trajectories that are constant in amplitude with changing principle axis are very fine-tuned, involving contrived changes in populations of all three domains (Fig.~\ref{fig:SI_microdomains}b). While this scenario is technically feasible, the strict restriction on the phase space has seemingly no physical basis. In particular, since we apply strain along $[210]$ ($\unit{30}{\degree})$ and $[100]$ ($\unit{0}{\degree})$, we would expect to change $x_0^+$ and $x_2^+$ symmetrically in the first case, and $x_1^+$ and $x_2^+$ symmetrically in the second. Clearly, this is not compatible with the trajectories in Fig.~\ref{fig:SI_microdomains}b. For example, to explain the rotation from $\Phi = \unit[30]{\degree}$ to about $\Phi = \unit[90]{\degree}$ at p1 (Fig.~\ref{fig:strain_2}b), $x_1^+$ and $x_2^+$ must at first have an opposite response to strain.

Furthermore, as shown in Fig.~\ref{fig:SI_histogram}, the amplitude has a relatively narrow distribution across the free sample. Therefore, we can use the same set of trajectories to understand how the detwinning scenario might reproduce the correlations between MCD and $\phi_0$ in both the free and strained sample. Since $x^+_n$ can be varied independently of $x^-_n$, it is possible at each angle to construct a distribution $\mathbf{x^-}$ that leads to the observed correlation. However, this clearly requires much more fine-tuning than the alternative explanation, where \Lvec varies continuously. On these grounds, we discard the detwinning scenario as a viable explanation for our observations.

\section{Minimal model for \Neel vector strain response}\label{sec:SM_strainmodel}

\begin{figure}
    \centering
    \includegraphics[scale=1]{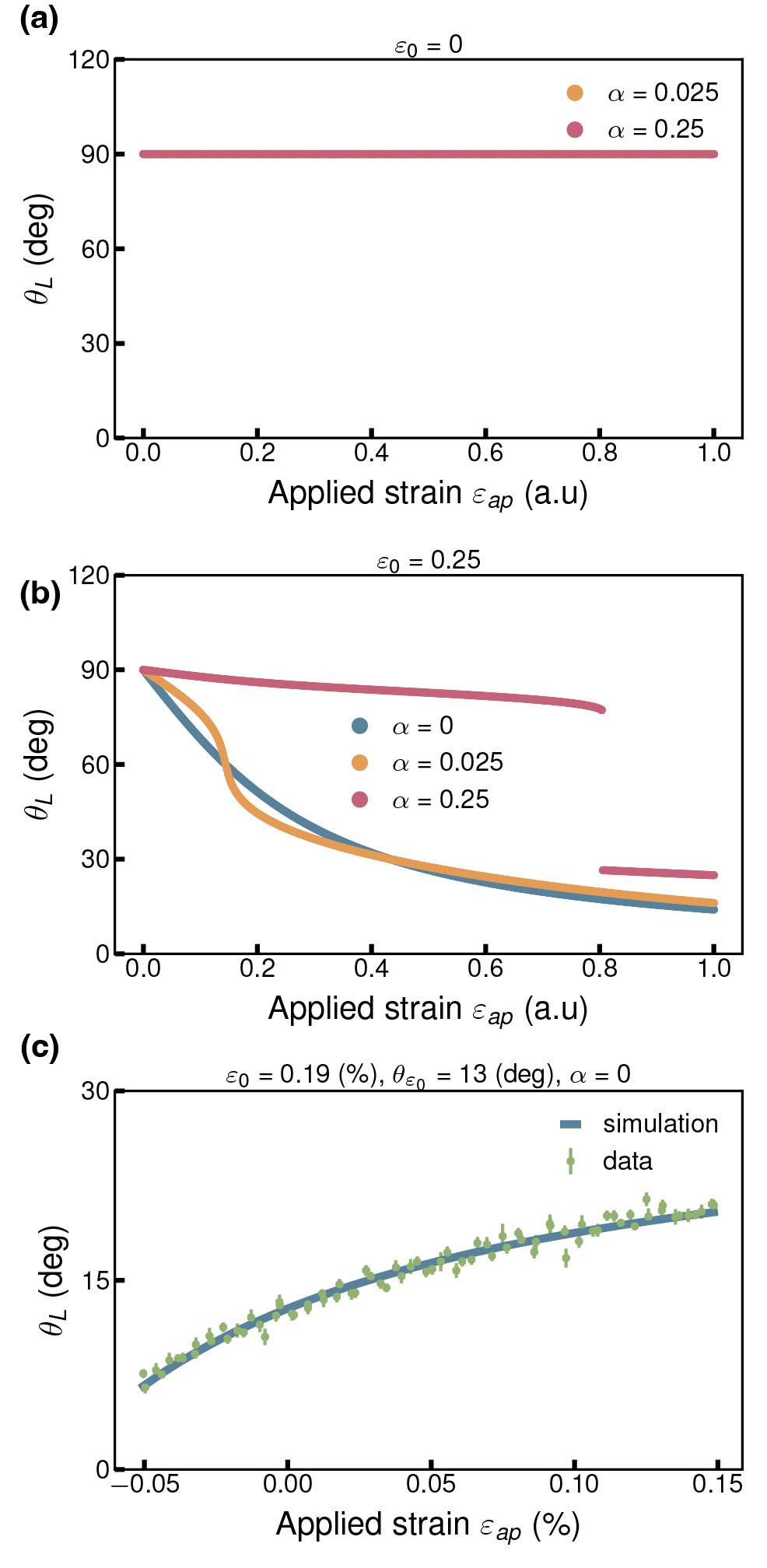}
    \caption{\textbf{(a-b)} Simulations for strain evolution of $\theta_L$ for applied strain along $\theta_{\varepsilon_{ap}} = \unit[0]{\degree}$, and MCA strengths $\alpha\in\{0, 0.025, 0.25\}$. Two cases are shown: (a) no built-in strain ($\varepsilon_0 = 0$), and (b) built-in strain $\varepsilon_0 = 0.25$ along $\theta_{\varepsilon_0} = \unit[90]{\degree}$. Smooth rotation of $\theta_L$ with applied strain is only observed for weak MCA in the presence of built-in strain. \textbf{(c)} Simulation fit to data shown in Fig.~\ref{fig:strain_1}d for $\alpha=0$ and $\theta_{\varepsilon_{ap}} = \unit[30]{\degree}$.}
    \label{fig:SI_strainmodel}
\end{figure}

Here, we construct a phenomenological model to better understand how strain can rotate \Lvec continuously, as shown in Fig.~\ref{fig:strain_1}. A minimal free energy for the \Neel vector orientation $\theta_L$ is
\begin{equation}
    F(\theta_L) = \alpha L^6\cos6\theta_L - \varepsilon L^2\cos2(\theta_L - \theta_\varepsilon),
    \label{eq:SM_free_energy}
\end{equation}
where the first term is the six-fold magnetocrystalline anisotropy (MCA), and the second term is the two-fold magnetoelastic (ME) anisotropy. MCA couples at sixth order in $L$, whereas ME couples at second order, indicative of the scaling with spin-orbit coupling (SOC). Since SOC is a small parameter, the ME can dominate over MCA in hexagonal systems. Here, we take $L=1$. The total strain is $\varepsilon$, and it is oriented at $\theta_\varepsilon$. For $\alpha>1$ the MCA favors $\theta_L = \unit{30}{\degree} + n\unit{60}{\degree}$, whereas $\epsilon>0$ favors $\theta_L = \theta_\varepsilon$ or $\theta_\varepsilon + \unit{180}{\degree}$.

We consider two scenarios for the strain evolution of $\theta_L$ at $T=0$. In Fig.~\ref{fig:SI_strainmodel} we show simulations of the strain evolution of $\theta_L$ in each case, obtained by local minimization of Eq.~\eqref{eq:SM_free_energy}. In the first case, we suppose that no built-in strain is present, and we study the strain evolution for $\alpha=\in\{0.025, 0.25\}$ and $\theta_\varepsilon = \unit{0}{\degree}$. $\theta_L$ starts at $\unit{90}{\degree}$, at an MCA minimum. The applied strain is taken from $\varepsilon = \varepsilon_{ap} = 0 \rightarrow 1$. Although the strain weakens the energy minimum at $\theta_L = \unit{90}{\degree}$, it does not change the orientation at which the minimum occurs, and therefore strain does not rotate $\theta_L$ (Fig.~\ref{fig:SI_strainmodel}a). For $\varepsilon > \alpha$, $\theta_L$ get stuck at an unstable energy maximum. At $T>0$, we expect that $\theta_L$ will jump from $\unit[0]{\degree}$ to $\unit[90]{\degree}$, but it will not do so continuously.

In the second case, in addition to the MCA, we include a built-in strain. The total strain on the sample is a combination of built-in strain $\varepsilon_0$ oriented at $\theta_{\varepsilon_0}$, and applied strain $\varepsilon_{ap}$ oriented at $\theta_{\varepsilon_{ap}}$. In particular, for the built-in strain at $\theta_{\varepsilon_0} = \unit{90}{\degree}$ so as to  align to the MCA easy axis, and applied strain at $\theta_{\varepsilon_{ap}} = \unit[0]{\degree}$, the total strain is
\begin{equation}
    \mathbf{\varepsilon}  = \varepsilon_0\hat{y} + \varepsilon_{ap}\hat{x}.
\end{equation}
We now consider three cases for the MCA: no MCA ($\alpha=0$), weak MCA ($\alpha = 0.025$), and MCA comparable to built-in strain ($\alpha = 0.25$). When $\varepsilon_{ap} = 0$, $\theta_L$ is oriented at $\unit{90}{\degree}$ to line up with the built-in strain and MCA. When the MCA is absent or weak compared to built-in strain, $\theta_L$ rotates smoothly towards $\unit{0}{\degree}$ with the applied strain (Fig.~\ref{fig:SI_strainmodel}a). Although $\theta_L$ also rotates when MCA is of the same strength as built-in strain, it does not rotate over the large range observed in experiments, and jumps are still present. This result, combined with the experiments shown in Fig.~\ref{fig:strain_1}, indicate that strain dominates over MCA, and that the orientation of $\theta_L$ is largely determined by the competition of built-in strain and applied strain. In Fig.~\ref{fig:SI_strainmodel}c, we show a simulation that reproduces the strain dependence of $\phi_0$ shown in Fig.~\ref{fig:strain_1}d. Here, $\alpha=0$, $\theta_{\varepsilon_{ap}} = \unit[30]{\degree}$, and the built-in strain parameters $\varepsilon_0 = 0.19$, $\theta_{\varepsilon_0} = \unit[13]{\degree}$ were obtained through a least-squares fit to the data. 

\section{Elastic lattice responses to strain}\label{sec:SM_elastic}

\begin{figure}
    \centering
    \includegraphics[scale=1]{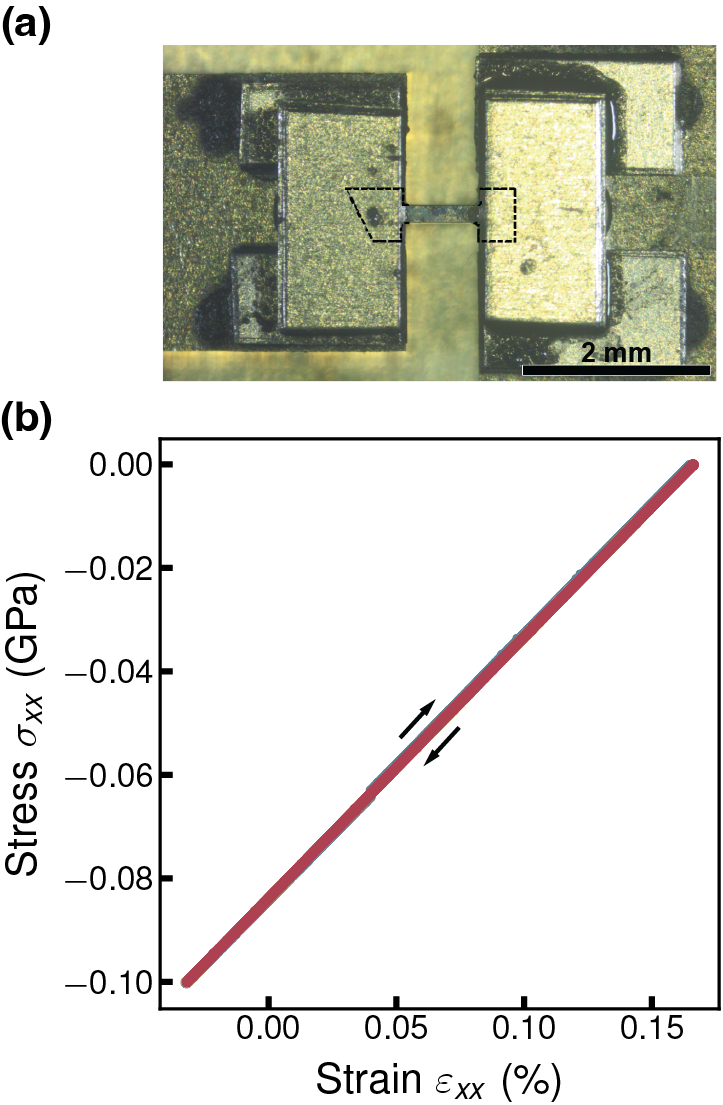}
    \caption{\textbf{(a)} Setup for measuring stress-strain curve in MnTe. A bar was cut from the same crystal as S2 and oriented with $[100]$ along the strain axis. \textbf{(b)}Elastic stress-strain response of MnTe at $\unit[25]{K}$ under similar cooling protocol to that of S2.}
    \label{fig:SI_elastic}
\end{figure}

In the main text, we observed hysteresis of the orientation of \Lvec with respect to applied strain (Fig.~\ref{fig:strain_2}a--b). We suggested that the hysteresis comes from plastic deformation of the magnetic subsystem. In that experiment, the sample was cooled down in an applied strain given by the differential thermal expansion between the titanium strain cell and MnTe, such that the sample was under about $\unit[0.23]{\%}$ strain to begin with. Another possible explanation for the hysteresis is that the ionic lattice itself underwent plastic deformation. In that case, if \Lvec were actually sensitive to stress, then one might expect to see hysteresis for a strain measurement. 

To check for mechanical hysteresis in MnTe, we performed stress-strain measurements on a third sample at $\unit[25]{K}$ using a home-built piezoelectric uniaxial stress cell with capacitive sensors for both force and displacement \cite{barber_piezoelectric-based_2019}. The sample was cut from the same crystal as S2, with the long axis along $[100]$. Following the methods described in \cite{noad_giant_2023}, we shaped the sample using a Xe plasma focused ion beam to create a well-defined mechanical system, then loaded it into the stress cell using Ti foils and Stycast 2850 epoxy. The dimensions of the exposed, narrow central portion, visible in Fig.~\ref{fig:SI_elastic}a, are $l \times w \times t = \unit[0.748]{mm} \times \unit[0.190]{mm} \times \unit[0.232]{mm}$. 

We cooled the stress-strain sample under comparable conditions to S2, without compensating for the differential thermal expansion. Once at $\unit[25]{K}$, we collected force and displacement capacitances CF and CD while slowly ramping the piezo control voltages until reaching a target value of CF. We then ramped in the reverse direction until reaching the original, as-cooled value of CF. We calculated the applied stress $\sigma$ by $\sigma = \Delta F/A$ where $F$ is the net force on the sample relative to the as-cooled, $\unit[25]{K}$ starting value and $A$ is the cross-sectional area of the narrow portion of the sample. We calculated the sample strain by the same methods described in Sec.~\ref{sec:SM_strain}. Here, the strain transmission factor was set to $\eta = 0.5$, which corresponds to the different mounting methods used compared to the method used in the magneto-optical strain measurements. The target CF was chosen to place the sample in compression relative to the starting value at $\unit[25]{K}$. Fig.~\ref{fig:SI_elastic}b shows that the stress-strain response is purely elastic. As a consistency check, we repeated the compression cycle and confirmed that the two datasets matched one another. This indicates that the hysteresis seen in sample S2 originates primarily from the magnetic subsystem.

% We also note that from this measurement we find that $C_{xxxx} = \unit[50.6]{GPa}$, which is a bit smaller than the theoretical value quoted in Eq.~\eqref{eq:SI_stress_strain}.

\end{document}